\documentclass[12pt,a4paper]{article}
\usepackage{amsmath,mathrsfs,graphicx,subfigure}
\usepackage{bbm}
\usepackage{chapterbib}
\usepackage{dcolumn}

\numberwithin{equation}{section}

\renewcommand{\vec}[1]{\mathbf{#1}}

\newcommand{\half}{\frac{1}{2}}

\newcommand{\vep}{\varepsilon}

\newcommand{\Rl}{R_{\lambda}}
\newcommand{\cdim}{C_{\vep}}

\newcommand{\cdiminf}{C_{\vep,\infty}}

\newcommand{\beq}{\begin{equation}}
\newcommand{\eeq}{\end{equation}}
\newcommand{\bea}{\begin{eqnarray}}
\newcommand{\eea}{\end{eqnarray}}

\newcommand{\dd}{\partial}
\newcommand{\ddt}{\frac{\partial}{\partial t}}

\newcommand{\xt}{(\mathbf{x},t)}
\newcommand{\kt}{(\mathbf{k},t)}
\newcommand{\pkt}{(\mathbf{k'},t')}

\newcommand{\jt}{(\mathbf{j},t)}
\newcommand{\lt}{(\mathbf{k-j},t)}
\newcommand{\kptp}{(\mathbf{k'},t')}
\newcommand{\kpptpp}{(\mathbf{k''},t'')}

\newcommand{\ua}{u_\alpha}
\newcommand{\ub}{u_\beta}
\newcommand{\ug}{u_\gamma}

\newcommand{\Mk}{M_{\alpha\beta\gamma}(\mathbf{k})}

\newcommand{\av}[1]{\left\langle #1 \right\rangle}

\newcommand{\ov}{\overline}

\title{The infinite Reynolds number limit and the quasi-dissipative anomaly}
\author{W. D. McComb and S. R. Yoffe\footnote{SUPA Department of
Physics, University of Strathclyde, John Anderson Building, 107
Rottenrow East. Glasgow G4 0NG.},\\
SUPA School of Physics and Astronomy,\\
Peter Guthrie Tait Road, \\
University of Edinburgh,\\
EDINBURGH EH9 3JZ.\\
Email: wdm@ph.ed.ac.uk}

\begin{document}

\maketitle 
\thispagestyle{empty} 
\begin{abstract} 

From a critical re-examination of Onsager's (1949) pioneering paper, we
find that his analysis was at odds with those of other workers because,
unlike others, he did not actually take the limit of zero viscosity $\nu
\rightarrow 0$. Instead, he simply set $\nu$ equal to zero, which is not
the same thing. The final sentence of his paper, where he asserted that
the detailed conservation symmetry does not hold globally because of the
`infinite number of steps' is contradicted by the analyses of Batchelor
(1953) and Edwards (1965). Onsager's work conflated the concepts of
infinite Reynolds number limit (in the mathematical sense) with that of
the breakdown of the continuum approximation. This was inconsistent, as
the former is based on a continuum mechanics approach, which relies on
the concept of an infinitely divisible fluid continuum; while the latter
requires a physics approach, with recognition of the underlying
molecular structure. We further show that his basic arguments are not in
accord with experimental results (some of which were available at the
time) which indicate that the onset of the zero-viscosity limit occurs
at quite modest Reynolds numbers, where there is no possibility of the
continuum approximation breaking down. This is the \emph{physical}
infinite Reynolds limit, which corresponds to the onset of
scale-invariance. We conclude that attempts to make the inviscid Euler
equation dissipative, by imposing constraints on its Fourier
representation, amount at best to a \emph{quasi-dissipative}
interpretation of the increasing energy occupation of an infinite
wavenumber space, and should be distinguished from the physics of
\emph{viscous} dissipation in real fluid turbulence.

\end{abstract}

\newpage
\tableofcontents
\newpage

\section{Introduction} 

In recent years the pioneering paper by Onsager \cite{Onsager49} has
exercised a growing influence on certain areas of fundamental research
in turbulence. This is generally to do with the nature of dissipation
and its relationship to the fluid viscosity in the limit of infinite
Reynolds number (or, equivalently, of zero viscosity). The fact that
dissipation appears to exist independently of viscosity in this limit,
is often referred to as the \emph{dissipation anomaly}, and basically
this terminology seems to reflect an acceptance of Onsager's view of the
matter.

Of rather more concern, in this view of things the infinite Reynolds
number limit is also seen as being equivalent to the breakdown of the
continuum description which underpins the Navier-Stokes equation.
Paradoxically, the Euler equation is regarded as surviving this
catastrophe, despite depending on precisely the same underlying
assumptions about the continuity of the fluid. It is argued that, in
effect, this paradox is evaded by working with the Fourier
representation of the Euler equation. As a result, the Euler equation is
supposed to account for the dissipation by means of mechanisms which
remain mysterious, but which are assumed to arise from constraints on
its Fourier representation. 

The purpose of the present paper is to make a critical examination of
these ideas. In view of the potential for confusion, we begin by
formally defining the dissipation rate and also establishing our notation.

\subsection{The viscous dissipation}

For a Newtonian fluid, the dissipation rate $\hat{\varepsilon}$ is formally defined in
terms of the coefficient of kinematic viscosity $\nu$, thus:
\beq
\hat{\varepsilon} = \frac{\nu}{2}\sum_{\alpha,\beta}\left(\frac{\dd u_\alpha}{\dd x_\beta} + \frac{\dd
u_\beta}{\dd x_\alpha}\right)^2;
\label{dissdefn}
\eeq
see, for example, the book by Batchelor \cite{Batchelor67}. Note that we
use Greek letters to denote the usual Cartesian tensor indices, where
these take the values 1, 2 or 3, as appropriate for a three-dimensional
space. There should be no confusion with the conventional use of Greek
indices in Minkowski four-space for the same purpose.

For isotropic turbulence, we have that $\ua\xt \equiv \mathbf{u}\xt$ is a random variable with zero mean, and
hence in this case $\hat{\varepsilon}$ is the \emph{instantaneous} dissipation rate,
and is also a random variable. For a turbulent flow we introduce the
\emph{mean} dissipation rate $\varepsilon$, as: 
\beq
\varepsilon = \frac{\nu}{2}\sum_{\alpha,\beta}\av{\left(\frac{\dd u_\alpha}{\dd x_\beta} + \frac{\dd
u_\beta}{\dd x_\alpha}\right)^2},
\label{mean-dissdefn}
\eeq
where the angle brackets $\av{\dots}$ denote the operation of taking an
average. 

We will use the \emph{mean} dissipation rate, along with the other
averaged quantities, in our description of fluid turbulence; and we will
use  undecorated symbols such as $\vep$ (and, for example, $E(k)$) for
this purpose. It should be noted that this practice has long been usual
(see for example the books by Batchelor \cite{Batchelor71} and by Landau
and Lifshitz \cite{Landau59}). However, it is by no means universal and
many people use $\ov{\varepsilon}$ or $\av{\varepsilon}$ for the mean
dissipation (or even $\epsilon$ rather than $\varepsilon$), so it is
necessary to be quite specific about this. Note that hereafter, when we
use the term `dissipation', we shall \emph{always} be referring to the
\emph{mean dissipation rate}. This is of course quite usual in the
subject, but we should emphasise that we will \emph{never} use the
unqualified term to refer to the instantaneous dissipation.

\section{Onsager (1949)}

The paper by Onsager (herafter referred to as Onsager49) is in two
parts. The first part deals with the statistical mechanics of
two-dimensional vortices in an ideal (i.e. frictionless) fluid. Onsager
concluded this part by wondering: ``how soon will the vortices discover
that there are three dimensions?''. He then commented:
\begin{quote}
The latter question is important because in three dimensions a mechanism
for complete dissipation of all kinetic energy, even without the aid of
viscosity, is available.
\end{quote}

In this paper we shall argue that this statement is misleading because
it is untrue. By dissipation, in its usual sense, we mean the the
conversion of macroscopic kinetic energy into microscopic random motion
which is perceived as heat. This is not the same as the absorption of
macroscopic kinetic energy in the infinitely large wavenumber space
which exists if the viscosity is zero. In the present paper we
consistently  put the case that the limit of infinite Reynolds numbers
must be explored by a standard limiting procedure. In this regard, we
are concerned with the second part of Onsager49, which is entitled
`Turbulence'. 

Not surprisingly, in view of its date of publication, Onsager49
is rather old-fashioned in its notation and procedures. In making our
commentary on this work, we will employ a fully modern notation, with
the occasional qualification in order to indicate its equivalent in
Onsager49. We will also divide our commentary into subsections, which we
hope will be helpful to the reader. We note that equations (9) - (13) of
Onsager49 constitute a real-space sub-section (in effect) and the
Fourier wavenumber treatment begins with his equation (14). We shall
discuss the real-space subsection first, with emphasis on his equation
(11) which is nowadays often referred to as the \emph{Taylor dissipation
surrogate}. 

\subsection{The Taylor dissipation surrogate}

Equation (11) of Onsager49 may be rewritten; first as:
\beq
\vep = -\frac{dU^2}{dt} = \cdim \frac{U^{3}}{L},
\eeq
where $\vep$ is the mean dissipation rate (Onsager calls it $Q$), the
first term on the right hand side is the negative of the energy decay
rate and the second term on the right is an intermediate step where
$\cdim$ is a \emph{coefficient} which is a function of the Reynolds
number; and secondly as:
\beq
\lim_{\Rl \rightarrow \infty }\vep = -\frac{dU^2}{dt} = \cdiminf \frac{U^{3}}{L},
\eeq
where $\cdiminf$ is the asymptotic constant that $\cdim$ becomes in the
limit of large Reynolds numbers. Other symbols are: $U$ is the
root-mean-square velocity and $L$ is the integral length scale.

Strictly speaking, it is the second of our two equations which is
equivalent to Onsager's equation (11); because he stated that \emph{`the
Reynolds number must be sufficiently large'}. We should also mention
that we are considering the specific case of free decay. In 1949, the
concept of stirring forces had not yet been introduced to the study of
isotropic turbulence and so presumably Onsager saw no necessity to
mention that it was force free.

At this point we begin our deconstruction of the turbulence section of
Onsager49, by focusing on a specific quotation which encloses his
equation (10) and goes as follows:
\begin{quote}
Experience indicates that for large Reynolds numbers the over-all rate
of dissipation is completely determined by the intensity $U^2$ together
with the ``macroscale'' $L$ of the motion, and the viscosity plays no
primary role, except through the condition that the Reynolds number
\dots must be sufficiently large.
\end{quote}

This statement is misleading. The viscosity enters into the dissipation
under all circumstances, and at all Reynolds numbers, through equation
(\ref{mean-dissdefn}). In order to assess its behaviour at large
Reynolds numbers it is necessary to apply a proper limiting procedure to
this expression. That has to be done in wavenumber space. It was not
done by Onsager, and we will return to how others have handled this in a
subsequent section. 

For the moment we confine our attention to the expression $\cdim U^3/L$.
Although this has often been referred to as the `Taylor dissipation
surrogate', it is actually a surrogate for the rate of inertial transfer
and becomes equal to the dissipation rate when the Reynolds number is
large enough and $\cdim \rightarrow \cdiminf$. This corresponds to the
onset of scale invariance of the inertial transfer through
wavenumber\footnote{Apparently this was first noted by Obukhov (see page
200 of the book by Lesieur \cite{Lesieur08}) and has since become an
important concept in the theory of turbulence}, and one can only discuss
this further in the Fourier representation. Accordingly we defer such
discussion to a later stage. For the moment we mention that these facts
have been established by McComb \emph{et al.}
\cite{McComb10b,McComb15a}, from both analysis and numerical simulation.
We may also mention in passing that (as we shall see later) this
asymptotic behaviour is observed for $\Rl \sim 100$, which is quite a
large Reynolds number but is in no danger of imperilling the continuum
nature of the fluid!

\subsection{The Navier-Stokes equations (NSE) in wavenumber space}

In Onsager49, the NSE appear as equation (13).
Then a Fourier-series representation is introduced through equations
(14,14a), and equation (15) is the resulting discrete form of the NSE in
wavenumber space. Between equations (13) and (14), another interesting
comment arises, thus:
\begin{quote}
Before we can arrive at a completely self-contained theory we shall have
to determine somehow, from the laws of dynamics, a statistical
distribution in function space, and for the time being we do not know
enough about how to describe such distributions.
\end{quote}
This is less clear than it might be. One assumes that Onsager was
referring to the probability distribution, but it is not clear whether
he was concerned about the theoretical procedure for obtaining such a
distribution, or about the mathematical aspects of representing it as a
functional. In either case, both problems were solved by Edwards in 1964
\cite{Edwards64}, who obtained the turbulence probability distribution
as an operator-product expansion about a Gaussian. This work was
followed by others and a recent review (and extension of the Edwards
analysis to the two-time case) can be found in
the paper by McComb and Yoffe \cite{McComb17a}.

In introducing Fourier series, Onsager referred only to a finite volume
$V$. Later on it became usual to refer to the fluid as occupying a box
in the form of a cube with a side of length $L_{\mbox{\scriptsize
box}}$, with periodic boundary conditions being imposed.  If we take the
limit $L_{\mbox{\scriptsize box}} \to \infty$, then we may replace sums
over wave-vectors by integrals, according to
\begin{equation}
	\lim_{L_{\mbox{\scriptsize box}}\to \infty} \left(
\frac{2\pi}{L_{\mbox{\scriptsize box}}} \right)^3 \sum_k =
	\int \mbox{d}^3 k.
\end{equation}
In the usual way we obtain $u_{\alpha}(\mathbf{k}, t)$, the Fourier transform of
$u_{\alpha}(\mathbf{x}, t)$ from
\beq
u_{\alpha}(\mathbf{k},t) = \left(\frac{1}{2\pi}\right)^{3}\int d^{3}x \,
u_{\alpha}(\mathbf{x}, t)\exp(-i\mathbf{k}\cdot\mathbf{x});
\label{ftkx}
\eeq
while the Fourier transform pair is completed by
\beq
u_{\alpha}(\mathbf{x}, t) = \int d^{3}k\, u_{\alpha}(\mathbf{k},
t)\exp(i\mathbf{k}\cdot\mathbf{x}).
\label{ftxk}
\eeq
In the 1970s, it became usual to just go directly to the
Fourier-transformed NSE, and the solenoidal form of this is now well
known. For the velocity field $\ua\kt$ in wavenumber $(k)$ space (see
either \cite{McComb90a} or \cite{McComb14a}) we have the solenoidal NSE as:
\beq
\left(\ddt + \nu k^{2}\right) \ua\kt=\Mk\int
d^{3}j \, \ub\jt\ug\lt,
\label{nsek}
\eeq
where the inertial transfer operator ${\Mk}$ is given by
\beq
\Mk = (2i)^{-1}\left[k_{\beta}P_{\alpha\gamma}(\mathbf{k}) +
k_{\gamma}P_{\alpha\beta}(\mathbf{k})\right],
\label{mop}
\eeq
and $i=\sqrt{-1}$,
while the projector $P_{\alpha\beta}(\mathbf{k})$ is expressed in terms
of the Kronecker delta as
\beq
P_{\alpha\beta}(\mathbf{k}) =
\delta_{\alpha\beta}-\frac{k_{\alpha}k_{\beta}}{k^{2}}.
\label{proj} 
\eeq
Note that the use of the projector ensures that the velocity field
remains solenoidal. Note also that our equation (\ref{nsek}) is equivalent to
equation (15) in Onsager49.

\subsection{Moments and spectra}

The covariance of the fluctuating velocity field may be introduced as
\beq
C_{\alpha\beta}(\mathbf{k,k'};t,t')= \langle\ua\kt\ub\pkt\rangle,
\label{velcov}
\eeq
and for isotropic, homogeneous turbulence we may write this
as
\beq
\langle\ua\kt\ub(-\mathbf{k},t')\rangle=P_{\alpha\beta}(\mathbf{k})\delta(\mathbf{k+k'})C(k;t,t'), 
\label{isvvelcov}
\eeq
where we anticipate the effects of homogeneity in writing the left hand
side. As is usual, the angle brackets $\av{\dots}$ denote the operation
of taking an average. 

If we consider the case $t=t'$, then we may introduce the spectral
density function:
\beq
C(k,t)\equiv C(k;t,t),
\label{spectolen}
\eeq
and the energy spectrum:
\beq
E(k,t) = 4\pi k^{2}C(k,t).
\label{espect}
\eeq

Lastly, we may also introduce the third-order moment, as:
\beq
\langle \ua\kt\ub\kptp\ug\kpptpp \rangle =
\delta(\mathbf{k}+\mathbf{k'}+\mathbf{k''})
C_{\alpha\beta\gamma}(\mathbf{k},\mathbf{k'},\mathbf{k''};t,t',t'').
\label{vel3mom}
\eeq
This will be used to obtain the energy transfer spectrum when we consider
the Lin equation.

\subsection{The statistical equations}

We form an equation for the covariance $C(k;t,t')$ in the usual way.
Multiply each term in (\ref{nsek}) by $u_{\alpha}(\mathbf{-k},t')$ and
take the average, to obtain:
\beq
\left(\ddt+\nu
k^{2}\right)C\left(k;t,t'\right)=\tfrac{1}{2}\Mk\int
d^{3}j\langle\ub\jt\ug\lt\ua(\mathbf{-k},t')\rangle,
\label{eq:closure}
\eeq
where we have used equations (\ref{proj}) and (\ref{isvvelcov}), cancelled
the factor $\delta(\mathbf{k+k'})$ across, and invoked isotropy, along
with the property
$Tr\, P_{\alpha\beta}(\mathbf{k}) = 2$. We can also write this in the
compact form:
\beq
\label{eq:comp-closure}
\left(\ddt+\nu
k^{2}\right)C\left(k;t,t'\right) = J(k;t,t'),
\eeq
where $J(k;t,t')$ is just the right hand side of \eqref{eq:closure}. The problem
of expressing this in terms of the covariance is the well-known
statistical closure problem.

Also using (\ref{nsek}), we can obtain an equation describing the energy
balance. On the time-diagonal, the covariance equation takes the form
for isotropic turbulence:
\begin{eqnarray}
	\left( \frac{\partial}{\partial t} + 2 \nu k^2
	\right) C(k,t)  = 
	Re \left[M_{\alpha\beta\gamma} (\mathbf{k}) \int \mbox{d}^3j
	\av{u_\beta(\mathbf{j},t)u_\gamma(\vec{k-j},t)u_\alpha(\vec{-k},t)} \right] +  F(k), 
\label{balance_deriv_f}
\end{eqnarray}
where $F(k)$ is the energy injection spectral density, as defined in
terms of the stirring forces, if present. 
Again, we can write this in a more compact form as:
\beq
\label{eq:comp-bal}
\left(\ddt+2\nu
k^{2}\right)C\left(k;t,t\right) = H(k;t,t) + F(k),
\eeq
where $H(k;t,t)$ can be determined by comparison with the right hand
side of (\ref{balance_deriv_f}). 

The energy spectrum is related to the spectral density by equation
(\ref{espect}). Accordingly, if we multiply (\ref{balance_deriv_f})
across by $4\pi k^2$, and rearrange terms, the governing equation of the
energy spectrum takes the form:

\begin{equation}
\frac{\partial E(k,t)}{\partial t} = W(k) + T(k,t)- 2 \nu k^2E(k,t),
\label{eq:lin}
\end{equation}
where the energy transfer spectrum is $T(k,t) = 4\pi k^2 H(k;t,t) \equiv   4\pi k^2
H(k,t)$. This equation for the energy spectrum is well known, and is
nowadays inreasingly referred to as the \emph{Lin equation}
\cite{Sagaut08,McComb14a}. For later convenience, we may also rearrange
it as:
\begin{equation}
-T(k) = I(k) - 2 \nu k^2E(k),
\label{figlin}
\end{equation}
where $I(k) = \partial E(k,t)/\partial t$ or $W(k)$, according
to whether we are studying free decay or forced, stationary isotropic
turbulence. Obviously in the former case, one could put in the explicit
time-dependences again.

\begin{figure} 
\begin{center}
\includegraphics[width=0.75\textwidth, trim=0px 200px 0px 200px,clip]{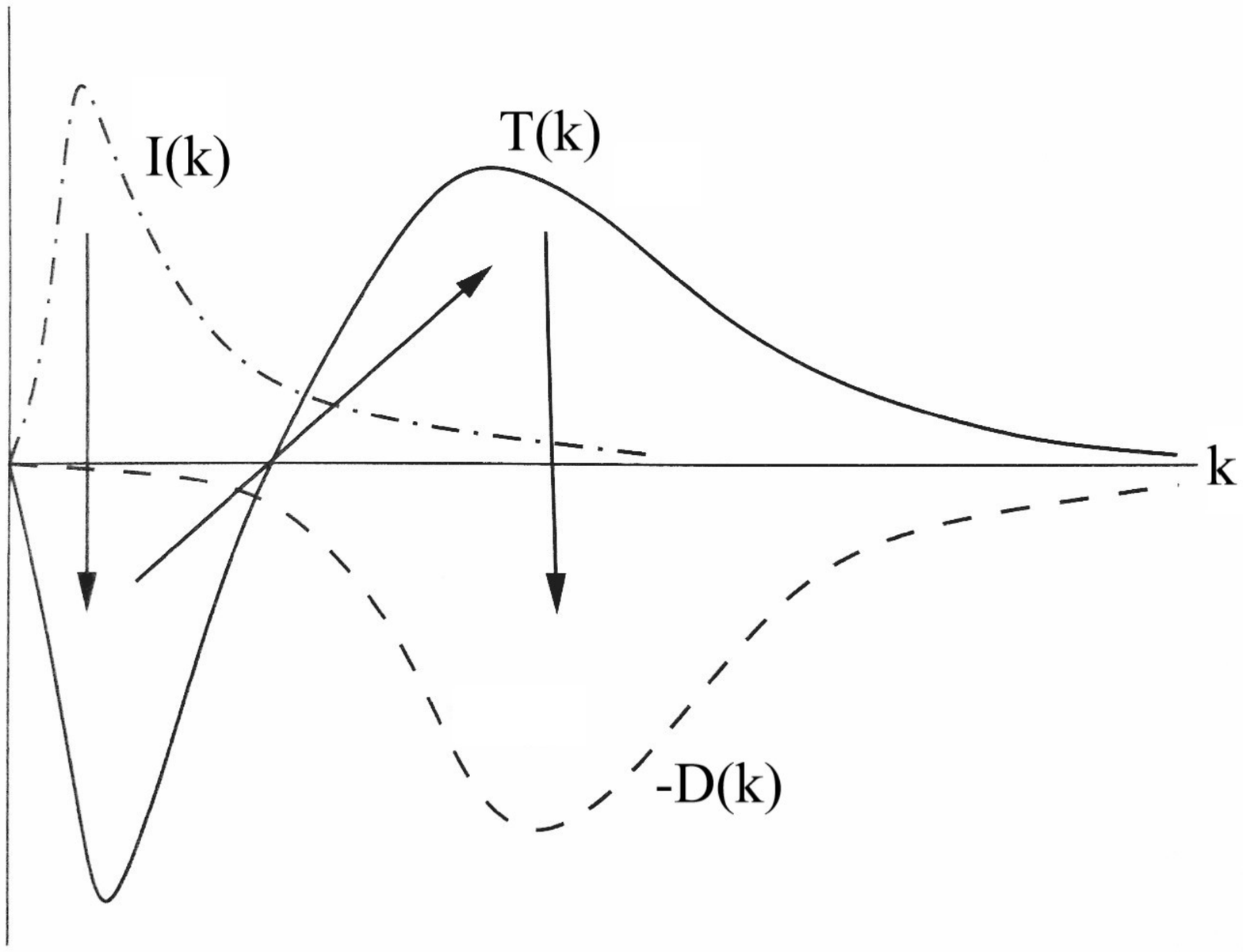}
\end{center} 
\caption{\small Schematic view of the terms in the Lin equation. The
input spectrum I(k) represents either the work spectrum $W(k)$ or $-\dd E(k,t)/ \dd
t$. The other symbols are as defined in equation (\ref{eq:lin}).} 
\label{linfig} 
\end{figure}

More detailed discussions of these equations
can be found in Chapter 3 of the book \cite{McComb14a}. A schematic view
of the energy transfer processes involved may be found in Fig.
\ref{linfig}, where we have used the notation of equation
(\ref{figlin}).

\subsection{Conservation properties of the inertial-transfer term}

In order to demonstrate the conservation properties of the energy
transfer spectrum $T(k)$, we make use of the triple-moment as defined by
equation (\ref{vel3mom}). Introducing the function $S(k,j;t)$ such
that:
\beq
T(k,t) = \int_0^\infty \, dj\, S(k,j;t),
\label{defskj}
\eeq
we have
\beq
S(k,j;t) = 2 \pi k^2j^2\int \, d\Omega_j
M_{\alpha\beta\gamma}(\mathbf{k})
\left[C_{\beta\gamma\alpha}(\mathbf{j},\mathbf{k-j},-\mathbf{k};t)
- C_{\beta\gamma\alpha}(-\mathbf{j},\mathbf{-k+j},\mathbf{k};t) \right],
\label{skjexp}
\eeq
where the solid angle $\Omega_j$ is defined from:
\beq
\int\,d^3j = \int_0^\infty\,j^2\,dj\int\,d\Omega_j.
\eeq
Not that interchanging $k$ and $j$ on the right hand side of equation
(\ref{skjexp}) maps the second term into the first, and vice versa,
hence the sign of right hand side changes and we conclude:
\beq
S(k,j;t) = - S(j,k;t).
\label{antisymm}
\eeq
Therefore it follows from (\ref{defskj}) that the integral of the transfer
spectrum over all wavenumber space vanishes, and hence energy is
conserved. A fuller treatment can be found in Chapter 3 of the book
\cite{McComb14a}.

Now Onsager's equivalent of our $S(k,j)$ is $Q(k,k')$ and his equation
(17) is just our equation (\ref{antisymm}), written is a slightly
different form. The point which we cannot emphasise strongly enough is
that this antisymmetry guarantees the conservation of turbulence kinetic
energy by nonlinear transfer for both the NSE and the Euler equations.
We shall return to this point later. First, we turn to how others
considered the actual limiting procedure which is involved in taking the
infinite Reynolds number limit.

\section{The infinite Reynolds number limit of Batchelor and others}

The earliest work on turbulence in the first part of the twentieth
century was on shear flows. Even then, systematic experiments with
increasing values of the Reynolds number led to an appreciation of a
limiting form of behaviour at large Reynolds numbers. In particular, the
work of Nikuradse in 1932 on pipe flows showed the development of
limiting behaviour of mean velocity profiles for the range $4 \times
10^3 \leq Re \leq 3.2 \times 10^6$. For further details and references
see the book by Goldstein \cite{Goldstein38}. A concise discussion can
also be found in Section 1.6 of \cite{McComb90a}. In addition, there was
much interest in the \emph{eddy-viscosity} concept, in which the
randomizing motions of turbulent eddies were interpreted as being analogous to
the randomizing motions of the molecules of the fluid. Such eddy
viscosities depended on the type of flow, but were generally found to be
something like two orders of magnitude larger than the molecular viscosity.  

Of course this corresponded to a much larger dissipation rate than in
laminar flow (usually interpreted as greater resistance to flow, in
those days) but it should be emphasised that this dissipation was not a
purely nonlinear effect. It was in fact a true dissipation of turbulent
kinetic energy due to the effects of the fluid viscosity. This was where
the cascade came in, as being a way of transferring turbulent energy to
scales where it would be most effectively turned into heat. We shall
come back to this point later, but for the moment we turn to the idea of
an infinite Reynolds number limit.

In his book on turbulence, (\cite{Batchelor71}: first published
in 1953), Batchelor introduced a local (in wavenumber) Reynolds number
for homogeneous turbulence and considered its behaviour as the viscosity
tended to zero. If we denote this by $R(k,t)$, then he defined it as:
\beq
R(k,t) = \frac{\left[E(k,t) \right]^{\half}}{\nu k^{\half}},
\label{local}
\eeq
where $E(k,t)$ is the energy spectrum, and noted that the effect of
decreasing $\nu$ is to increase the dominance of the inertia forces over
the viscous forces for the motion associated with that degree of
freedom. He went on to say:
\begin{quote} 
Consequently the region of wave-number space which is affected
significantly by the action of viscous forces moves out from the origin
towards $k=\infty$ as the Reynolds number increases. \emph{In the limit
of infinite Reynolds numbers the sink of energy is displaced to infinity
and the influence of viscous forces is negligible for wave-numbers of
finite magnitude.} 
\end{quote}
Note that the emphasis is ours. 

A more extensive discussion of the infinite Reynolds number limit can be
found in Section 6.2 of Batchelor's book (\emph{ibid}), but the fragment
given here is sufficient for our present purposes. This idea was
extended by Edwards \cite{Edwards65}, who came to the same conclusion
from a consideration of the Kolmogorov dissipation wavenumber in the
limit of infinite Reynolds number, and represented the dissipation rate
by a delta function at infinity. In order to study the stationary case,
he balanced this with an input term in the form of a delta function at
the origin. 

Of course, if we interpret `infinity' in the purely mathematical sense,
then the above limiting cases are only valid for a picture of fluid
motion which is part of continuum mechanics. In practice a real fluid
has a molecular structure which limits the continuum approximation to
length scales larger than inter-atomic spacings. To take account of this
it is better to use the concept of the inertial flux of energy and this
we will now do.

\subsection{Scale-invariance of the inertial flux as the infinite
Reynolds number limit}

The concept of the inertial flux (and its scale-invariance) was already
known to Obukhov in 1941 (see reference \cite{Lesieur08}), used by
Onsager in 1945 \cite{Onsager45}, and discussed by Batchelor in 1953
\cite{Batchelor53}. But its first formal use was by Kraichnan
\cite{Kraichnan59b} who used it to test the inertial-range behaviour of
his direct-interaction approximation. In the process, he introduced a
useful symbol for this flux, which he referred to as the \emph{transport
power} and defined (actually in the later form to be found in
\cite{Kraichnan64b}) as
\beq
\Pi(\kappa,t) = \int_\kappa^\infty  dk\, T(k,t) = - \int_0^\kappa  dk
\, T(k,t),
\label{tp}
\eeq
where $\Pi(\kappa,t)$ is the flux of energy through wavenumber $\kappa$
due to the nonlinear term.

It was understood at an early stage that the inertial flux would
increase with increasing Reynolds number until it became equal to the
dissipation rate $\varepsilon$; and that this was the largest value that
the flux could achieve. Increasing the Reynolds number further merely
increased the extent of wavenumber space, as the Kolmogorov dissipation
wavenumber was also increased. Accordingly, it was recognised that
the condition:
\beq
\Pi = \varepsilon,
\label{criterion}
\eeq
was the criterion for the onset of the inertial range of wavenumbers.
This was stated in words in the original (1953) edition of Batchelor's book
\cite{Batchelor71} and in the lectures by Saffman \cite{Saffman68}. It
can be found in the books by Leslie \cite{Leslie73}, McComb
\cite{McComb90a,McComb14a}, Davidson \cite{Davidson04}, Lesieur
\cite{Lesieur08}, Sagaut and Cambon \cite{Sagaut08}, and Verma \cite{Verma19}. The criterion
is used in direct numerical simulation (e.g. see references
\cite{McComb01}, \cite{Ishihara09} and \cite{McComb14b}) and in
theoretical work (e.g. see the papers by Chasnov \cite{Chasnov91},
Bowman \cite{Bowman96}, Qian \cite{Qian97},
Thacker \cite{Thacker97}, Falkovich \cite{Falkovich06}, and Lundgren
\cite{Lundgren08}), where the fact that the criterion holds over a range
of wavenumbers is usually referred to as \emph{scale-invariance}.

The onset of scale-invariance may be taken as equivalent to attaining
the infinite Reynolds number limit. As the Reynolds number is increased
further, there can be no resulting qualitative change. The inertial
range of waveumbers, with its constant flux of energy through it, merely
becomes more extensive. In order to understand this better, it may be
helpful to consider how the transfer spectrum $T(k)$ behaves under these
circumstances and this is the subject of the next sub-section.

\subsection{Dependence of the transfer spectrum $T(k)$ on Reynolds
number}

The first measurements of the transfer spectrum were obtained by Uberoi
\cite{Uberoi63} in 1963 for freely decaying turbulence. Uberoi obtained
the shape of $T(k)$, as shown schematically in Fig. \ref{linfig}, by
measuring the decay rate of the energy spectrum and the dissipation; and
using the Lin equation (in the form given by equation (\ref{figlin})) to
evaluate the transfer spectrum. A direct measurement of $T(k)$ was first
carried out in  1969 by Van Atta and Chen \cite{Vanatta69}, and of
course nowadays DNS and closure approximations produce this curve
routinely.

As an aside, we should mention that Uberoi found his results rather
surprising, in that he expected that the scale-invariance condition $\Pi
= \varepsilon$ would be paralleled by the condition $T=0$, for the same
range of wavenumbers. The fact that the transfer spectrum crossed the
$k$-axis at a single point he attributed to the Reynolds number being
too small to show the effect properly. This led Lumley to introduce an
approximate criterion \cite{Lumley64} to allow the inertial range to be
identified using the transfer spectrum; and an example of its use can be
found in \cite{McComb92}. Later more extensive investigations confirmed
that $T(k)$ possessed only a single zero-crossing \cite{Bradshaw67},
\cite{Helland77}. Ultimately, in 2008 this puzzle was formulated as a
paradox and resolved by McComb \cite{McComb08}, who introduced a
decomposition of the transfer function, based on the antisymmetry of the
exchange function $S(k,j)$. More recently a modified Lin equation was
introduced in order to avoid the paradox in the first place
\cite{McComb20}. However, we mention these matters mainly for sake of
completeness and we will not pursue them further here. What follows is
based mainly on the more recent book by McComb \cite{McComb14a}, but we
should also mention the comprehensive analysis of Tchoufag, Sagaut and
Cambon \cite{Tchoufag12}, which deals with both real-space and spectral
representations. 

\begin{figure} 
\begin{center}
\includegraphics[width=0.75\textwidth, trim=0px 10px 0px 0px,clip]{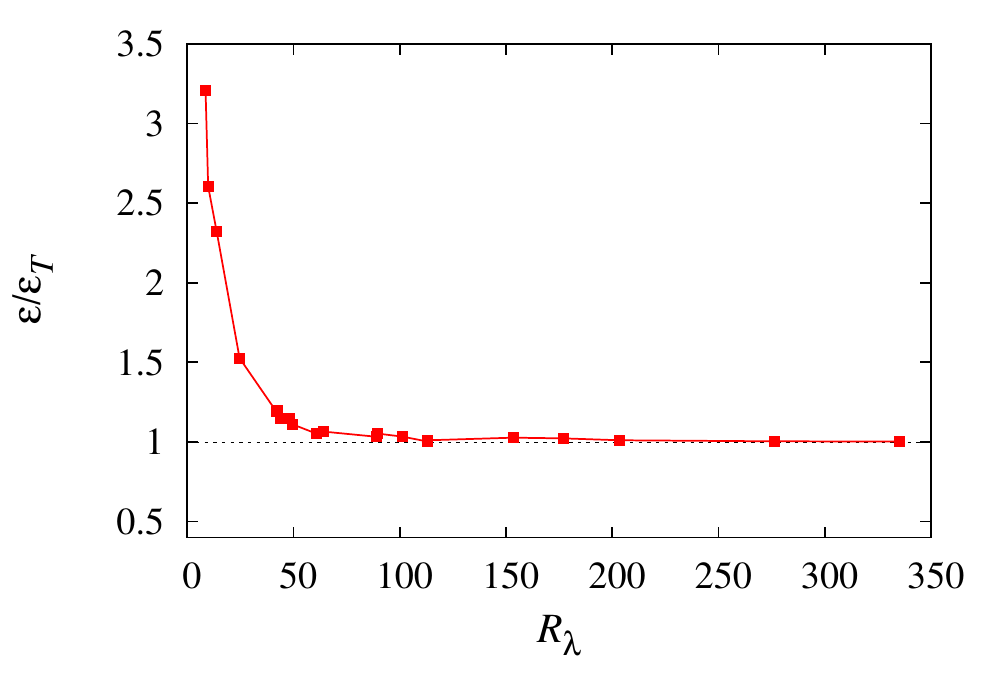}
\end{center} 
\caption{\small  Variation of the mean dissipation rate $\varepsilon$
for forced turbulence, divided by the
peak inertial transfer rate $\varepsilon_T \equiv \Pi_{max}$, with increasing Taylor-Reynolds number,
showing the onset of scale-invariance and hence the limit of infinite
Reynolds numbers. Data are openly available online from the University of Strathclyde KnowledgeBase
[http://dx.doi.org/10.15129/64a4a042-7d0d-48ce-8afa-21f9883d1e84].} 
\label{infre} 
\end{figure}

We can approach the idea of an infinite Reynolds number limit, by
considering the maximum value of the flux which must correspond to the
point where the transfer spectrum becomes zero. We will denote this
value of the wavenumber by $k_\ast$, and hence we may write:
\beq
T(k_\ast) = 0 \qquad \mbox{and hence}\qquad \Pi_{max} = \Pi(k_\ast).
\eeq
$\Pi_{max}$ is the maximum value of the flux at any Reynolds number, but
as the Reynolds number increases, it is bounded from above by the
dissipation rate. Thus, introducing the symbol $\varepsilon_T =
\Pi_{max}$, we have the criterion for having reached the infinite Reynolds
number limit:
\beq
\mbox{Criterion for the onset of infinite Reynolds number behaviour:} \quad \frac{\Pi_{max}}{\varepsilon}
= 1
; \quad \mbox{or} \quad \frac{\varepsilon_T}{\varepsilon}=1.
\eeq
In Fig. \ref{infre} we show the development of the infinite Reynolds
number limit. In fact we plot the reciprocal of the above criterion, as
this figure is taken from an investigation that was actually studying
the behaviour of the dissipation rate \cite{Yoffe12}. Nevertheless, the
qualitative behaviour should be perfectly clear. Those readers
interested in seeing the details of how this behaviour develops in the
spectral picture should consult reference \cite{Shanmugasundaram92}.

It is of interest to reconcile the infinite Reynolds number limit of
Edwards \cite{Edwards65} with the present treatment. We may rewrite
equation (\ref{figlin}) for the stationary case, and then take the
infinite Reynolds number limit, thus:
\begin{equation}
-T(k) = W(k) - 2 \nu k^2E(k) = \varepsilon_W \delta(k) -\varepsilon \delta(k-\infty),
\label{sfe}
\end{equation}
where we note that energy conservation gives us $\varepsilon_W =
\varepsilon$.

Of course, the Edwards result only applies (as does Batchelor's) in
continuum mechanics. But it can be useful if one is testing a statistical
closure theory which is also applied to a fluid in the continuum
mechanics description. This is where the transport power comes in. From
equation (\ref{tp}) applied to (\ref{sfe}), we have:
\beq
\Pi(\kappa) = \varepsilon_W = \varepsilon.
\eeq
It may seem that ({\ref{sfe}}) is in conflict with the experimental
finding that the transfer spectrum has a single zero-crossing. It is of
interest to note that Tchoufag \emph{et al.} \cite{Tchoufag12} found
that, when the input and output regions were well separated, $T(k)$ took
the form of a \emph{quasi-plateau}, being a linear region which was at a
small angle to the axis. As the Reynolds number increases, this angle
becomes smaller, and evidently the Edwards result would be appropriate
for a true continuum fluid at where the Reynolds number is actually
infinite. But of course the use of the criterion based on the flux
brings it into conformity with the more physical picture.

\section{The nature of viscous dissipation}

In fluid dynamics, the term `dissipation' invariably means `viscous
dissipation'. Moreover, the term `viscosity' is short for `coefficient
of viscosity', and assigns a magnitude to a process in which relative
motion of a fluid is randomized and converted to molecular motion; which
in turn corresponds to a rise in the temperature of the fluid. In other
words, viscous dissipation is a thermodynamic process; although normally
the thermodynamic aspects can be neglected for all practical purposes. 

From this it follows that, contrary to Onsager49, when the viscosity is
zero there is no viscous dissipation, and hence no dissipation at all.
Evidently, with an infinitely large wavenumber space, energy can be
absorbed and endlessly transferred to ever increasing wavenumbers. But
this disappearance of turbulent kinetic energy can at best be described
as \emph{quasi dissipation}. 

With reference to the last sentence in Onsager49, it is worth pointing
out that Euler's equation has been used to study absolute equilibrium
ensembles in fluid dynamics. It is well known that these systems, when
subject to an arbitrary initial random field, have equipartition
solutions. Of course such systems have to be truncated to finite volumes
of wavenumber space; but there is no reason to suppose that, as the size
of the space is increased, there is any deviation from this
behaviour. For a general treatment, see the book by Lesieur
\cite{Lesieur08}; for some specfic remarks relevant to our present
discussion see the paper by Kraichnan \cite{Kraichnan64b}; and
for some other applications see \cite{Kraichnan73,Kraichnan80}.

When the viscous term is added back in (i.e. restoring the NSE), the
effect is symmetry-breaking, because of the factor $k^2$. So the
viscosity restores the uni-directional energy transfer process and also,
through damping the higher-order modes, removes the need for truncation.
This brings us back to the question of the supposed breakdown of the
continuum description at high Reynolds numbers. This was raised as a
question by Batchelor (see page 5 of \cite{Batchelor71}), who answered
his own question as follows:
\begin{quote}
However, the action of viscosity is to suppress strongly the small-scale
components of the turbulence and we shall see that for all practical
conditions the spectral distribution of energy dies away effectively to
zero long before length scales comparable with the mean free path are
reached. As a consequence we can ignore the molecular structure of the
medium and regard it as a continuous fluid.
\end{quote}
Leslie went further and presented an order-of-magnitude calculation in
his book (see pages 3-4 of \cite{Leslie73}), from which he concluded that, at
a Reynolds number of $10^6$ in pipe flow, the smallest turbulence scales
would be three orders of magnitude greater than the inter-molecular
spacing of the fluid.

This calculation is unusual and perhaps unique. It is curious how often
workers in the field express concern about possible singularities, or
advocate some course of action to circumvent them, but never actually
present even an estimate of the conditions under which they may be
expected to occur.

\section{Conclusion}

We have concluded that, with the modern interpretation of the infinite
Reynolds limit as being equivalent to the onset of scale-invariant
inertial transfer, there is really no need to suppose that it amounts to
the viscosity actually being equal to zero. Nor is there any reason to
suppose that the simple symmetry $S(k,j) = -S(j,k)$, which guarantees
both energy conservation in the NSE and equipartition in the Euler
equation, will mysteriously disappear under any of the limiting
processes available to us. Accordingly, we are unable to see what
constitutes an anomaly. The Fourier representation of the Euler equation
cannot be altered to destroy the above symmetry in any physically
meaningful way. The Euler equation is inviscid and conservative and does
not possess the ability to dissipate energy in the usual sense of real
fluid mechanics. 

The term anomaly is often used in connection with the observed limiting
behaviour when the dimensionless dissipation is plotted against Reynolds
number. But this process is fully understood at a spectral level, so
there does not seem to be any justification of this usage. However, if
the term is used to mean that the Euler equation can somehow be made to
give the appearance of dissipating energy, then the most we would
concede is that this could be described as a quasi-dissipation anomaly.

Lastly, in this work we have been concerned with the different ways in which the concept of infinity arises in mathematics and in physics. This topic has recently been receiving some attention in the more philosophical reaches of both subjects: see the paper by van Wierst \cite{Wierst19} and references therein. This author sees the situation in the theory of critical phenomena as paradoxical, and suggests that the paradox can be resolved by resorting to constructive mathematics. However, we would argue that the paradox can be resolved by the recognition that the concept of infinity in the relevant branches of physics (including the present work) can be replaced by the onset of scale invariance. This is a matter that we intend to pursue further.

\section*{Acknowledgement}

SRY acknowledges support from the UK EPSRC (grant number EP/N028694/1)
and from the STFC (grant number ST/G008248/1)


\end{document}